\documentclass[letterpaper,aps,prd,preprint,amssymb,showpacs,nofootinbib,superscriptaddress]{revtex4}

\usepackage{amssymb}
\usepackage{graphicx}

\newcommand{\nc}{\newcommand}
\nc{\postscript}[2] 
{\setlength{\epsfxsize}{#2\hsize}\centerline{\epsfbox{#1}}}
\nc{\non}{\nonumber}
\nc{\hc}{\hbox {h.c.}} \nc{\re}{\hbox {Re}} 
\nc{\mev}{\hbox {MeV}} \nc{\gev}{\;\hbox {GeV}} \nc{\tev}{\;\hbox {TeV}}
\def\lsim{\mathrel{\raise.3ex\hbox{$<$\kern-.75em\lower1ex\hbox{$\sim$}}}}
\def\gsim{\mathrel{\raise.3ex\hbox{$>$\kern-.75em\lower1ex\hbox{$\sim$}}}}

\nc{\etal}{{\it et al.}}
\nc{\Lsp}{\;\;\;\;\;\;\;\;\;\;}  \nc{\LLLsp}{\lspace \lspace}
\nc{\lsp}{\;\;\;\;\;\;}
\nc{\spac}{\;\;\;}
\nc{\noi}{\noindent}
\nc{\beq}{\begin{equation}}   \nc{\eeq}{\end{equation}}
\nc{\bea}{\begin{eqnarray}}   \nc{\eea}{\end{eqnarray}}
\nc{\baa}{\begin{array}}      \nc{\eaa}{\end{array}}
\nc{\bit}{\begin{itemize}}    \nc{\eit}{\end{itemize}}
\nc{\ben}{\begin{enumerate}}  \nc{\een}{\end{enumerate}}
\nc{\bce}{\begin{center}}     \nc{\ece}{\end{center}}

\def\lcal{{\cal L}}
\def\sq2{\sqrt{2}}

\def\ph{\varphi}

\def\m4{m^4(\ph)}
\def\mn2{m_n^2}

\def\v5{V^{(5)}}
\def\subA{{\!{}_A}}
\def\subAl{{\!{}_{A,\ell}}}

\begin{document}

\title{\begin{flushright}
\vspace{-.2cm}
       \mbox{\normalsize \rm SU-4252-865}

\vspace{-.35cm}
       \mbox{\normalsize \rm UMD-PP-07-009}
       \end{flushright}
       \vskip 20pt
Existence and Stability of Non-Trivial Scalar Field Configurations in Orbifolded Extra Dimensions}
\author{Manuel Toharia\footnote{mtoharia@physics.syr.edu}} 
\affiliation{Department of Physics, Syracuse University\\ Syracuse NY 13244, USA}
\affiliation{Department of Physics, University of Maryland\\ College Park MD 20742 USA}
\author{Mark Trodden\footnote{trodden@physics.syr.edu}}
\affiliation{Department of Physics, Syracuse University\\ Syracuse NY 13244, USA}

\date{\today}

\begin{abstract}
We consider the existence and stability of static configurations of a scalar field in a five dimensional spacetime in which the
extra spatial dimension is compactified on an $S^1/Z_2$ orbifold. For a wide class of potentials with multiple minima there
exist a finite number of such configurations, with total number depending on the size of the orbifold interval. However, a
Sturm-Liouville stability analysis demonstrates that all such configurations with nodes in the interval are unstable. 
Nodeless static solutions, of which there may be more than one for a given potential, are far more interesting, and we present
and prove a powerful general criterion that allows a simple determination of which of these nodeless
solutions are stable. We demonstrate our general results by specializing to a number of specific examples, one of which may
be analyzed entirely analytically.
\end{abstract}

\maketitle

\section{Introduction}
\label{intro}
The possibility of extra spatial
dimensions~\cite{Kaluza:tu,Klein:tv,Rubakov:1983bb,Akama:1982jy,Antoniadis:1990ew,Lykken:1996fj,
Arkani-Hamed:1998rs,Antoniadis:1998ig,Randall:1999ee,Randall:1999vf,Lykken:1999nb,
Arkani-Hamed:1999hk,Antoniadis:1993jp,Dienes:1998vg,Kaloper:2000jb,Cremades:2002dh,Kokorelis:2002qi}, hidden from our current
experiments and observations through compactification or warping, has opened
up a wealth of options for particle physics model building and allowed
entirely new approaches for addressing cosmological problems.

In many models, standard model fields are supposed to be confined to a
submanifold, or brane, while in other models they populate the entire
bulk. Common to both approaches, however, is the inclusion of bulk fields
beyond pure gravity, either because they are demanded by a more complete
theory, such as string theory, or because they are necessary to stabilize the
extra dimensional manifold. Thus, a complete understanding of the predictions and allowed
phenomenology of extra dimension models necessarily includes a comprehensive
consideration of the configurations of these fields. 

The allowed configurations of such bulk fields are determined, naturally, by their equations of motion, 
subject to the boundary conditions imposed by the particular extra-dimensional model under
consideration. These might be periodic boundary conditions, in the case of a
smooth manifold, or reflection-symmetric ones in the case of an orbifolded
extra dimension.

In this paper, building on our recent letter~\cite{ttletter} we concern ourselves with a class of allowed nontrivial 
scalar field configurations~\cite{
Arkani-Hamed:1999dc,Georgi:2000wb,Kaplan:2001ga,Manton:1988az,Sakamoto:1999yk,Hung:2003cj,Grzadkowski:2004mg} in
orbifolded extra-dimensional models, neglecting gravity. These
configurations exist whenever the potential possesses
at least two degenerate minima
and we show that they may form a finite tower of kink state solutions. We explicitly demonstrate
that all but the lowest-lying of these - the ones with no nodes in the
interval - are unstable. In addition we identify a general stability criterion for these
lowest-lying states, and provide concrete examples for specific convenient
choices of potential.

That a finite tower of 
nontrivial static configurations may exist, with the possibility of
multiple stable ones, allows for new phenomena and
constraints on the models, and may have wide-ranging
implications for the particle physics and cosmological theories constructed
around
them~\cite{Arkani-Hamed:1998nn,Macesanu:2004gf,Starkman:2001xu,Starkman:2000dy,Deffayet:2001xs,
Deffayet:2001pu,Dvali:2000hr,Deffayet:2002sp,Deffayet:2000uy,Dvali:2003rk,Lue:2004rj,Lue:2002sw,
Binetruy:1999hy,Binetruy:2001tc,Binetruy:1999ut,Chung:1999zs,Csaki:1999mp,Cline:2002ht}.

We are currently considering the effects of including gravitational effects on
the configurations explored in this
paper.

\section{General Scalar Potential}

Since we are neglecting gravity for the entirety of this paper, our background is a flat $4+1$ dimensional spacetime, 
with coordinates ${x^M\equiv (x^{\mu},y)}$, with indices $M,N,\ldots
=0,1,2,3,5$, $\mu,\nu,\ldots =0,1,2,3$. The extra dimension $x^5 \equiv y$ is
compactified on an orbifold $S_1/Z_2$ defined by $y\in (0,\pi R)$, with size
$\pi R$ assumed to be fixed.

Propagating on this background, we consider a real scalar field defined by the action
\beq
S=\int d^5x\, \left[{1\over2} \eta^{MN} \partial_M \phi(x,y) \;\partial_N \phi(x,y) - V (\phi)\right] \ .
\label{scal_models_general}
\eeq
Because of the orbifolded geometry, we can demand that the scalar
field $\phi(x,y)$ be odd under $Z_2$ reflections along the extra coordinate
(i.e. $\phi(x,y)=-\phi(x,-y)$). 

To ensure this, we require that the potential $V(\phi)$ be invariant under the discrete symmetry
$\phi\to-\phi$ and, to simplify notation, we also choose the potential to vanish at $\phi=0$.
We will be particularly interested in potentials which possess multiple degenerate minima, the simplest examples of which are
those with two degenerate global minima at $\phi=\pm v$ with $v\neq 0$. 

\subsection{Properties of Static Solutions}

We seek static field configurations $\phi_{{}_A}(y)$, parametrized by their amplitudes $A$,
which extremize the action, and with nontrivial $y$-dependence, subject to the appropriate boundary conditions,
namely $\phi_{{}_A}(0)\!=0\!$ and $\phi_{{}_A}(\pi R)\!=\!0$. 

\begin{figure}[h]
\begin{center}
\includegraphics[width = 9cm]{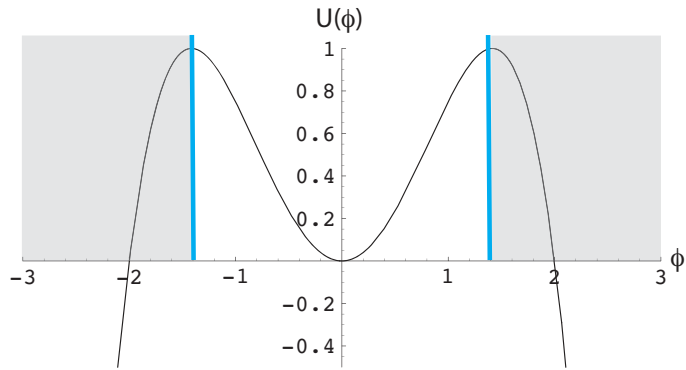}
\end{center}
\vspace{-.5cm}
\caption{Mechanical Analogy: Periodic solutions of a particle in the potential
  $U(\phi)=-V(\phi)=(\mu^2/2)\phi^2-(\lambda/4)\phi^4$ (here with $\mu^2=2$ and $\lambda=1$) exist when the
  total energy of the particle lies between $E_{\rm max}={\mu^4\over 4\lambda}$ (top of the inverted potential) and
  $E_{min}=0$. A particle with energy $E_{{}_A}$ will undergo a
  periodic motion of period $T$, understood as the length of the extra-dimension. Note that this is precisely the potential 
  used in our first example~(\ref{v5}).}
\label{mechanicalanalog}
\vspace{.0cm}
\end{figure}

The field equation satisfied by such solutions is
\bea
\phi_{{}_A}{\!\!\!''}- {\partial V \over \partial \phi_{{}_A}} = 0 \ ,
\label{scal_el_general}
\eea
where a prime denotes a derivative with respect to $y$.
It is easily seen that there exists a first integral, given by 
\beq
\frac12 \phi_{{}_A}{\!\!'}^2 + U(\phi_{{}_A})=E_A \ .
\label{energy_generalE}
\eeq
where $U(\phi)=-V(\phi)$ and $E_A$ is a constant. This choice of nomenclature will be convenient for much of this
paper, since it is helpful to think of this problem as that of the
position $\phi(y)$ of a particle rolling in time $y$ without friction in the 
inverted potential $U(\phi)$ \cite{Coleman:1977py} (see Figure~\ref{mechanicalanalog}). 

Since the potential $U(\phi)$ vanishes at $\phi=0$
and the solution $\phi_\subA(0)$ also vanishes at $y=0$, by evaluating~(\ref{energy_generalE}) 
at $y=0$ we see that $E$ is determined by the value of the kink derivative
$\phi'_\subA(y)$ at $y=0$ via $2E=\phi_\subA^{'\;2}(0)$ (in the mechanical
analogy, $E$ is the total energy of the system, which at $y=0$ is all kinetic energy).

We may rewrite~(\ref{energy_generalE}) as
\beq
\frac12 \phi_\subA^{'\;2}-V(\phi_\subA)=-V(A) \ ,
\label{energy_generalVA}
\eeq
in which we use the fact that the total energy of the system is equal to the
potential energy evaluated at the point where the magnitude of the background
solution attains its maximum value, its amplitude $A$.

Through the mechanical analogy, it is relatively straightforward to see that periodic solutions
can only exist for $A<\phi_{min}$, where $\phi_{min}$ is the global maximum of
$U(\phi)$ (or the global minimum of $V(\phi)$).

Since the amplitude $A$ parametrizes the different possible nontrivial solutions, it will prove useful to write
$\phi_\subA\equiv\phi_\subA(y,A)$.

In this notation, we may write the period $T(A)$ of the solution $\phi_\subA(y,A)$ as
\bea
T(A)=2 \sqrt{2} \int_0^A {dX\over \sqrt{V(X)-V(A)}} \ ,
\label{TA}
\eea 
which must be related to the radius $R$ of the extra dimension.

As noted in~\cite{Grzadkowski:2004mg}, the physical 
size $\pi R$ of the extra dimension does not need to be equal to the half
period $T/2$ of the background solution $\phi_\subA(y)$, but rather must be a multiple of it
\beq
2\pi R=(\ell\!+\!1)\ T \ ,
\label{periodradius}
\eeq
with $\ell=0,1,..,\ell_{\rm max}$ an integer. Solution(s) with $\ell=0$ will be
nodeless in the interval $(0,\pi R)$, while solutions with $\ell>0$ will have $\ell$
nodes between the two boundaries of the orbifold $0$ and $\pi R$.

Any solution $\phi_\subA(y)$ that vanishes at the two fixed points of the
orbifold $y\!=\!0$ and $y\!=\!\pi R$, 
must have vanishing derivative $\phi'_\subA(y)$ at at least one
intermediate value of $y$.

The symmetry of the potential implies that the points $y\!=\!{m_\ell\ T \over
  4(\ell+1)}$, where $m_\ell=1,..,2\ell+1$, are always
such special points i.e.
\beq
\phi'_\subA\left[\frac{m_\ell\ T}{4 (\ell+1)}\right]=0 \ .
\eeq
At all these points, the magnitude of the background solution attains its
maximum value $A$.

Let us now assume that the radius $R$ of the extra dimension is 
fixed\footnote{We will assume that some mechanism fixes and stabilizes $R$
without affecting anything else in the setup.}. As we have already mentioned
and we shall see, there are, in general, multiple
nontrivial background solutions corresponding to a given radius, with the precise number depending on the specific
choices of the potential $V(\phi)$ and on the radius $R$.  We will identify two classes of solutions among these; background
solutions with nodes, and those that are nodeless.  

\subsection{Enumerating Solutions}

The physical size of the extra dimension is related to the period of the background 
solution $\phi_\subA(y)$ by~(\ref{periodradius}).

To see that the maximum number of nodes $\ell_{\rm max}$ is finite and
provides a lower bound on the maximum number of independent nontrivial
solutions, consider the period function $T(A)$ and focus once again on the
mechanical analogy. 
When the potential $U(\phi)$ has a local minimum at $\phi=0$ (as in Figure \ref{mechanicalanalog}),
it is clear that in the limit $A\to 0$ the period of a nontrivial solution to Eq.~(\ref{energy_generalE}) will
be determined purely by the quadratic part of the potential $U(\phi)$,
i.e. $T(0)={2\pi\over\mu^2}$, where $\mu^2={\partial^2 U\over \partial
  \phi^2}\Big|_{{}_{\phi=0}}$. This is because $\phi$ must remain small in that
limit and one can neglect higher order terms in the potential $U(\phi)$
leaving only the quadratic term. In this limit the system becomes a simple
harmonic oscillator, with frequency set by the quadratic coefficient of
the potential. 

As the amplitude $A$ is increased, the period $T(A)$ may increase or decrease,
but can never decrease to zero (since the ``time'' it takes to complete a period
can never be zero).

When the potential $U(\phi)$ has a local maximum at $\phi=0$, nontrivial
solutions with an amplitude $A\to 0$ do not exist. There will be a minimum
value of $A$ for which nontrivial solutions exist, and the period $T(A)$ will diverge at that value.

Note also that, if we allow $A$ to approach the value of $\phi$ at a different
local maximum, the period $T(A)$ once again diverges.

Thus, in all cases there exists a global minimum of $T(A)$ that we denote by $T_{\rm min}$.
For fixed $R$, nontrivial solutions exist only if $T_{\rm min}\leq 2\pi R$. 
It follows that in a size $2 \pi R$ there exist at least $\ell_{\rm max}$ nontrivial
solutions, where 
\beq
\ell_{\rm max}=IP\left(\frac{2\pi R}{T_{\rm min}}\right)-1
\eeq
and $IP(x)\equiv$ IntegerPart$(x)$ gives the largest integer less than or equal to $x$.

A nontrivial solution $\phi_\subAl(y)$ contains $\ell$ nodes in the orbifold
interval. Note that we have now used the number of
nodes $\ell$ along with the amplitude $A$ to parametrize the solutions.

It is, as we shall see explicitly later, possible that there exist two or more solutions with different
amplitudes, $A_1$ and $A_2$, say, but with the same period $T(A_1)=T(A_2)$.
In particular, since, as we have argued, the function $T(A)$ has a global minimum, if the physical size of the extra dimension is
$2\pi R=T(A_1)=T(A_2)$ there exist two nodeless nontrivial kink solutions to our problem.


\section{Stability of Nontrivial Solutions}
For a given potential $V(\phi)$ and a given size $\pi R$ of the orbifold interval, we have shown how to enumerate
and construct all possible nontrivial static configurations of our scalar field $\phi$. The existence of these
configurations is somewhat interesting in its own right, but their physical relevance will depend on their stability properties.

To study this, we begin by adding small perturbations around a given solution $\phi_\subAl(y)$ 
of~(\ref{scal_el_general}), writing
\beq 
\phi(x,y) = \phi_\subAl(y) +\varphi(x,y) \ ,
\eeq
where we are again parametrizing the background solution with its amplitude $A$ and
its number of nodes $\ell$. The 5D Lagrangian then becomes, up to terms quadratic 
in the perturbations
\bea
\lcal^{(5)} &=& \lcal^{(5)}_\subAl +\frac12 \; \partial^\mu \varphi(x,y) \;\partial_\mu
\varphi(x,y) - \frac12 \varphi(x,y) \left[-{d^2\over dy^2}+{\partial^2 V\over \partial
    \phi^2} \Big|_{\phi_\subAl} \right] \varphi(x,y)+..\ \
\label{}
\eea
where $\lcal^{(5)}_\subAl$ is the lagrangian density corresponding to the background solution.

From this Lagrangian we may obtain the equations of motion of the field
$\varphi(x,y)$. Writing $\varphi(x,y)=\varphi_x(x) \varphi_y(y)$ these become
\bea
\Box \varphi^n_x(x) &=& -M_n^2 \varphi^n_x(x)\\
-{\varphi_y^n}^{''}(y) + q(y) \varphi^n_y(y) &=& M_n^2\ \varphi^n_y(y)
\label{kkmodes}
\eea
where
\beq
q(y)={\partial^2 V\over \partial\phi^2} \Big|_{\phi_\subAl} \ .
\eeq

These are the equations of motion of a tower of 4-dimensional scalar fields $\varphi^n_x(x)$
with squared masses $M_n^2$ and with extra-dimensional profile functions $\varphi^n_y(y)$, which
are determined by solving~(\ref{kkmodes}).


\subsection{Instability of Solutions with Nodes in the Interval $(0,\pi R)$}

A useful result for dealing with those solutions with nodes is obtained by
taking the derivative of the equation for the background
solution~(\ref{scal_el_general}), yielding
\bea
\phi_\subAl'''(y) - \left({\partial^2 V\over \partial
    \phi^2} \Big|_{\phi_\subAl(y)}\right) \phi_\subAl'(y)=0 \ .
\label{kinkder}
\eea

Comparing Equations~(\ref{kinkder}) and~(\ref{kkmodes}) we see that the
derivative $\phi_\subAl'(y)$ of the background solution can be identified as a massless 
($M_n^2=0$) solution to~(\ref{kkmodes}),
{\it but with Neumann boundary conditions rather than the Dirichlet ones we require}
\footnote{This should not come as
a big surprise, since it is just the translation mode, the masslessness of
which is a reflection of translation symmetry\cite{Rajaraman,Lee:1991ax}. It cannot be a
physical solution since translation invariance is broken in the orbifold.}.

We now appeal to the general theory of eigenvalues of the Sturm-Liouville problem with Dirichlet (D), 
periodic (P), semiperiodic (S), and Neumann (N) boundary conditions. That theory contains the following chain of inequalities
\bea
\lambda_0^N &\leq& \lambda_0^P <\lambda_0^S \leq \{\lambda_0^D,\ \lambda_1^N\}\leq \lambda_1^S <\lambda_1^P
\leq \{\lambda_1^D,\ \lambda_2^N\} \nonumber \\
&\leq& \lambda_2^P <\lambda_2^S \leq \{\lambda_2^D,\ \lambda_3^N\}\leq \lambda_3^S <\lambda_3^P
\leq \{\lambda_3^D,\ \lambda_4^N\} \nonumber \\
&\leq& \cdots \ ,
\label{eigenvalueinqual}
\eea
relating the towers of eigenvalues corresponding to each different
eigensolution $\varphi_i^D$, $\varphi_i^P$, $\varphi_i^N$ and $\varphi_i^S$
defined by each type of boundary condition.

Applying this to any scalar configuration $\phi_\subAl (y)$ with greater than the minimal
periodicity ($\ell\!>\!0$), we see that the associated derivative $\phi_\subAl'(y)$, obeying
Neumann boundary conditions, will have multiple nodes 
in the interval $(0,\pi R)$. Thus we may identify it as the eigensolution
$\varphi_i^N(y)$, with $i\geq 2$, with its masslessness (from comparing~(\ref{kinkder}) and~(\ref{kkmodes}))
implying that the corresponding eigenvalue obeys $\lambda_i^N =0$. 

However~(\ref{eigenvalueinqual}) implies that
$\lambda_2^N>\lambda_0^D$. Therefore, if $\lambda_i^N=0$ for some $i\geq 2$,
then there exists at least one ($\lambda_0^D$) eigenvalue of the Dirichlet
problem, and possibly more, that are negative!

Thus, {\it all static solutions with nodes in the interval are unstable}.


\subsection{Stability of Nodeless Solutions}

We now turn our attention to the study of perturbations around a solution $\phi_\subA(y)$ with no nodes in
the interval $\ (0,\pi R\!=\!T/2)$ (and therefore parametrized only by the
amplitude $A$). We focus on the sign of the eigenvalue $\lambda$ of the lowest eigenfunction 
of equation~(\ref{kkmodes}), which we rewrite as
\beq
\varphi''(y)- \left[q(y)-\lambda\right] \varphi(y) =0 \ ,
\label{STeq}
\eeq 
where $\lambda= M_0^2$, and $\varphi(y)$ is the lowest lying eigensolution,
which obeys the Dirichlet boundary conditions $\varphi(0)=\varphi(T/2)=0$.

An important step in our proof of stability will be the study of the massless scalar excitations. We have
already identified one such solution, $\varphi^N_1(y)=\phi_\subA'(y)$, the
derivative of the background profile, but it is not a physical one, since it
satisfies Neumann boundary conditions instead of Dirichlet ones. Nevertheless, this solution does allow us to
construct a second, linearly independent solution via
\beq
\varphi_2(y)=\varphi^N_1(y) \int_0^y {ds\over \varphi^N_1(s)^2} = \phi_\subA'(y)
\int_0^y {ds\over {\phi_\subA'}^2(s)}\label{phi2}
\eeq

This solution automatically satisfies a Dirichlet boundary condition at $y=0$, but to identify it 
as a physical solution, we need to establish the
circumstances under which it obeys such a condition at $y=T/2$.

In this regard, it is useful to note that equation~(\ref{STeq}), with
$\lambda=0$, is in the form of the {\it Hill equation}, for which the
following theorem (see, for example~\cite{Magnus}) holds

{\it \noindent Let $Y_1(t)$ and $Y_2(t)$ be two
differentiable solutions of the Hill equation 
\beq
Y''(t)+ Q(t)\ Y(t) =0 \ ,
\label{Hilleq}
\eeq
with $Q(t)=Q(t+T/2)$, uniquely determined by the conditions,
\bea
Y_1(0)=1,&& Y_1(0)'=0,\non\\
Y_2(0)=0,&& Y_2(0)'=1.
\eea
\noindent When $Q(t)\!=\!Q(-t)$ and when $Y_1'(T/2)\!=\!0$ and $Y_1(T/2)\!=\!-1$, 
then
\beq
Y_2(T/2)=0\ \Longleftrightarrow\ Y_2(T/4)'=0
\eeq
}
This means that, assuming that $Q(t)$ is even, and that the solution $Y_1(t)$ satisfies
$Y_1(0)=1,\ Y_1(T/2)=-1, \ Y_1(T/4)=0$ and its derivative $Y'_1(y)$ satisfies 
$Y_1'(0)=Y_1'(T/2)=0$, then $Y_2(t)$ will obey Dirichlet boundary conditions
at $t=0$ and $t=T/2$, if and only if it obeys a Dirichlet
boundary condition at $t=0$ and a Neumann one at $t=T/4$, 

This theorem applies precisely to our problem -- equation~(\ref{STeq}) with $\lambda=0$,
$Q(t)\equiv -q(y)$, and where the function $q(y)$ is even in $y$ due to the symmetry of the potential $V(\phi)$.
Thus, we infer that $\varphi_2(y)$ will be a physical solution if and
only if it obeys a Dirichlet condition at $y=0$ and a Neumann one at $y=T/4$. As it turns out, this
condition at $y=T/4$ is simpler to study than the one at $T/2$.

Our problem is therefore mapped to that of identifying parameter values for which $\varphi_2'(T/4)=0$.

Differentiating equation~(\ref{phi2}) gives
\beq
\varphi_2'(y)=\phi_\subA''(y)\int_0^y {ds\over {\phi_\subA'}^2(s)} + {1\over \phi_\subA'(y)} \ ,
\eeq
and so our condition for the existence of a massless scalar excitation is 
\beq
\phi_\subA''(T/4)\int_0^{T/4} {ds\over {\phi_\subA'}^2(s)} + {1\over \phi_\subA'(T/4)}=0 \ .
\eeq
The two terms separately formally diverge, but this divergence must cancel when
they are added together.

Now recall our expression~(\ref{TA}) for the period $T(A)$ of a solution as a
function of the amplitude $A$. Taking a derivative with respect to $A$ yields
\bea
{dT\over dA}&=& {\partial V(A)\over \partial A}
\int_0^A { \sqrt{2}\ dX \over (V(X)-V(A))^{3\over2}}+{2 \sqrt{2}\over \sqrt{V(X)\!-\!V(A)}}
{\left|\vphantom{\int^\int_{\int_{A}}}\right.}_{\!\!{}_{X\to
    A}},\ \
\label{dTdA}
\eea
and using~(\ref{scal_el_general}) and~(\ref{energy_generalVA}) we may rewrite this as
\bea
{dT\over dA}&=&4\left( {1\over \phi_\subA'(T/4)}+   \phi_\subA''(T/4) \int_0^{T/4}
  {dy\over {\phi'_\subA}^2(s)}\right) \ ,
\eea 
or simply
\bea 
{dT\over dA}= 4\ \varphi_2'(T/4) \ .
\label{varphi2primzero}
\eea

Thus, {\it a massless scalar excitation around a background solution of equation~(\ref{scal_el_general}), 
with amplitude $A_c$, exists if and only if the derivative $dT/dA$ of the period function $T(A)$
vanishes at $A=A_c$. Moreover, this
nodeless massless excitation will be the lowest eigenvalue solution of the problem.}

Let us now return to solutions with non-zero eigenvalues. The Rayleigh-Ritz variational result applied 
to~(\ref{STeq}) yields an expression for the eigenvalue $\lambda$ in terms of the 
eigenfunction $\varphi(y)$
\bea
\lambda= {\int_0^{T/2}\left(\varphi'(y)^2+q(y)\varphi(y)^2\right)\ dy \over\int_0^{T/2} \varphi(y)^2\ dy} \ .
\eea
Because the potential $q(y)$ satisfies $q(y+T/2)=q(y)$ and is symmetric around
$y=0$ and $y= T/4$, it is sufficient to consider the half interval $(0,T/4)$, since the eigenfunctions
will be either symmetric or antisymmetric around $y=T/4$.

Thus
\bea
\lambda= {2\over N} {\int_0^{T/4}\left(\varphi'(y)^2+q(y)\ \varphi(y)^2\right)\ dy } \ ,
\eea
where $N=2\int_0^{T/4} \varphi(y)^2\ dy$.

Now assuming that the eigenvalue $\lambda$ and the eigenfunction
$\varphi$ are differentiable with respect to the amplitude parameter $A$, one
can show that 
\bea
{\partial\lambda\over \partial A}\Big|_{A=A_c} &=& {2\over N}\int_0^{T/4}\varphi^2 {\partial
    q\over\partial A}\Big|_{A=A_c} \ ,
\label{dlamdAc}
\eea
where $A_c$ is such that $\left.{dT\over dA}\right|_{A=A_c}=0$ and therefore is an amplitude for which
the lightest scalar excitation around the kink solution is massless.  

The variation of $q(y)$ with $A$ can be written as 
\beq
{\partial q\over\partial A} = -{\partial V\over\partial A}
\left({\partial^2 V \over \partial \phi^2}\right)' I(y) \ ,
\eeq
where
\beq
I(y)= \int^y_0{ds\over {\phi'}^2(s)} \ .
\eeq
Thus
\beq
{\partial\lambda\over \partial A}\Big|_{A_c} =  -{2\over N}{\partial
  V\over\partial A}\ \beta_{A_c} \ ,
\eeq
where we have defined the integral $\beta_{A_c}$ 
\beq
\beta_{A_c}=\int_0^{T/4} \left({\partial^2 V \over \partial \phi^2}\right)' {\phi'}^2 I^3 dy \ ,
\eeq
which may be integrated by parts successively 
to give
\beq
\beta_{A_c}={\partial^2 V \over \partial A^2}{I(T/4)\over\left(\partial
    V\over\partial A\right)^2}
-{I(T/4)\over{\phi^{'}}^2(T/4)} +3\int_0^{T/4} {dy\over {\phi'}^4} \ .
\eeq
Putting all this together then yields
\beq
{\partial\lambda\over \partial A}\Big|_{A_c} = -{2\over N}{\partial V\over\partial A}
\left[
{\partial^2 V \over \partial A^2}{I(T/4)\over\left(\partial V\over\partial A\right)^2} 
-{I(T/4)\over{\phi^{'}}^2(T/4)}
+3\int_0^{T/4} {dy\over {\phi'}^4} 
\right] .
\label{dellam}
\eeq

To complete the proof, now consider the second derivative of the period function $T(A)$,
evaluated at $A=A_c$
\bea
{d^2T\over dA^2}\Big|_{A_c}&=& 4  {\partial^2 V\over\partial A^2}I(T/4)\ 
-\  4 \left({\partial V\over\partial A}\right)^2 {I(T/4)\over {\phi'}^2(T/4)}
+\ 12\left({\partial V\over\partial A}\right)^2 \int^{T/4}_0 {dy \over {\phi^{'}}^4} .
\eea
Comparing this to~(\ref{dellam}) evaluated at $A=A_c$, we obtain our final result
\beq
{\partial\lambda\over \partial A}\Big|_{A_c} = -{1\over 2N}{1\over {\partial
    V\over\partial A}}\ {d^2T\over dA^2}\Big|_{A_c} \ .
\eeq

This is our central result, and the proof of stability follows: 

\begin{itemize}
\item We have demonstrated that at points $A=A_c$ at which ${dT\over dA}$ vanishes, the lightest scalar excitation is
massless.
\item We have also proved that at $A=A_c$, since $-{\partial V\over\partial A}>0$, the sign of the derivative with 
respect to $A$ of the lightest eigenvalue is entirely determined by the sign of  ${d^2T\over dA^2}\Big|_{A_c}$. 
\item This means that for any nontrivial background solution of amplitude $A$, the sign of $\lambda$ will be the 
same as the sign of ${dT/dA}$. To see this, consider an interval $A \in
(A_{c_1},A_{c_2})$, over which $T(A)$ is a continuous function of $A$, and
where $A_{c_1}$ and $A_{c_2}$ are two consecutive critical values at which
$dT/dA$ vanishes, but at which $d^2T/dA^2$ is nonzero.

If $dT/dA>0$ inside that interval (except perhaps at points of inflection) then $d^2T/dA^2$
is strictly positive at $A_{c_1}$, and therefore $d\lambda/dA$ is also
strictly positive there. Thus in this case $\lambda$ is positive in the whole interval. 
If, on the other hand, $dT/dA<0$ inside the interval, then by an identical argument, $\lambda$ must also be negative. 

It remains to point out that, in the case in which there exists a single critical value of $A$ in a region over which $T(A)$ is
continuous, then the above argument still holds, but with the point $A_{c_2}$ replaced by the value of $A$ at which $T(A)$
becomes singular.
\end{itemize}

We have thus established our general stability criterion:
{\it A static,
  nodeless solution $\phi_{{}_{A_*}}(y)$ to equation~(\ref{scal_el_general}),
with amplitude $A_*$, and period $T(A_*)$, and satisfying
  $\phi_{{}_{A_*}}(0)=\phi_{{}_{A_*}}(T/2)=0$, is stable if and only if
\beq
\left.\frac{dT}{dA}\right|_{A=A_*} > 0 \ .
\label{stabilitycriterion}
\eeq
} 
Before moving on to some examples, it is worth discussing what happens when
we perturb around the trivial solution $\phi_\subA(y)=0$. We have seen that there
exists a minimal size $\pi R=T_{\rm min}/2$ of the orbifold interval for which one can
find nontrivial static solutions. When the size of the orbifold
is smaller than that critical size, the only static background solution
possible is the trivial one.

In this case~(\ref{kkmodes}) becomes
\beq
{\varphi^n}^{''}(y) - (\mu^2 - M_n^2)\ \varphi^n(y) = 0 \ ,
\label{kktrivial}
\eeq
where $\mu^2={\partial^2 V\over \partial\phi^2} \big|_{{}_{\phi=0}}$ may be
either positive or negative.

A general solution to this equation is
\beq
\varphi^n(y) = C\cos{\left(\sqrt{M_n^2-\mu^2}\right) y} +D
\sin{\left(\sqrt{M_n^2-\mu^2}\right) y} \ .
\eeq
Imposing the Dirichlet boundary condition at $y=0$ requires that $C=0$, and imposing the same at $y=\pi R$ yields
\beq
M_n^2={(n+1)^2\over R^2}+\mu^2 \ .
\eeq

The stability of the trivial solution depends on the sign of $M_0^2$ and is
guaranteed when $\mu^2>0$. However, even if $\mu^2<0$ the solution
will be stable as long as $|\mu^2|<{1\over R^2}$.

This result is quite interesting since it becomes clear now that the trivial solution must be treated
carefully given that it can be part of the group of stable static solutions to
equation~(\ref{scal_el_general}). The stability of the trivial
solution $\phi_\subA(y)=0$ depends on $R^2$ being smaller than $|{1\over\mu^2}|$, so as long as
the potential $V(\phi)$ allows the existence of nontrivial stable
kink solutions with period $T<{2\pi\over\mu^2}$, then these solutions will
coexist with the trivial solution as the complete set of static classically
stable configurations.

\section{Energy Density of Nontrivial Configurations}

Since, as we have shown, it is possible for there to exist multiple nodeless
and classically stable configurations, we would like to compute the energies
of each of these in order to determine the vacuum state of our extra dimensional scalar.
This is because quantum mechanical effects will make the highest energy configurations
metastable, eventually decaying into the lowest energy
configuration, which should then be treated as the true vacuum.

Making explicit once again the dependence on the amplitude parameter $A$, the energy of a static configuration is
\beq
E(A)=2 \int_0^{T_\subA/2} \left({1\over2}{\phi_\subA'}^2 +V(\phi_\subA) \right) dy \ ,
\eeq
where again, for simplicity, we assume that $V(0)=0$.

Using~(\ref{energy_generalVA}) this becomes
\bea
E(A)&=&T(A)V(A)+4\sqrt{2} \int_0^{A}\sqrt{V(\phi)-V(A)}\ d\phi \ ,
\label{EofA}
\eea
from which we obtain
\bea
{\partial E\over \partial A}&=& {\partial T\over \partial A}V(A) 
\label{dEbydA}
\eea 
where we have used the definition of $T(A)$ in~(\ref{TA}).

Now, $V(\phi)$ is always negative when evaluated at an amplitude $A$ at
which there exists a nontrivial solution. Also, when such as solution is
stable, we have already shown that ${\partial T\over \partial A} >0$. 

Therefore, over any range of $A$ for which $T(A)$ is a continuous function, 
${\partial E\over \partial A}<0$,  when evaluated on a stable nontrivial configuration. A possibly interesting
corollary of this result is that if one were to consider the equivalent solution on an interval of slightly
larger size, this would inevitably have a higher amplitude and therefore a lower energy density. Thus, the 
energy density of a given solution is lowered by making the interval larger. This may have important ramifications for the 
stabilization of extra dimensions, which we have not considered here, but are pursuing in other work.

\section{Examples}

After this rather general and formal treatment of the stability properties of static solutions, we now turn to
some concrete examples with which to better understand the results.

\subsection{Example 1: Mexican-Hat Potential}

Our first example is exactly solvable, and the existence of the relevant
solutions has been thoroughly studied in~\cite{Grzadkowski:2004mg}. Consider the potential
\beq
V_1(\phi)= -{\mu^2 \over 2} \phi^2 +{{\bar \lambda} \over 4} \phi^4\ ,
\label{v5}
\eeq
where $[\mu]=[{\bar \lambda}]^{-1}=({\rm Mass})$.  
\begin{figure}[h]
\begin{center}
\includegraphics[width = 9cm]{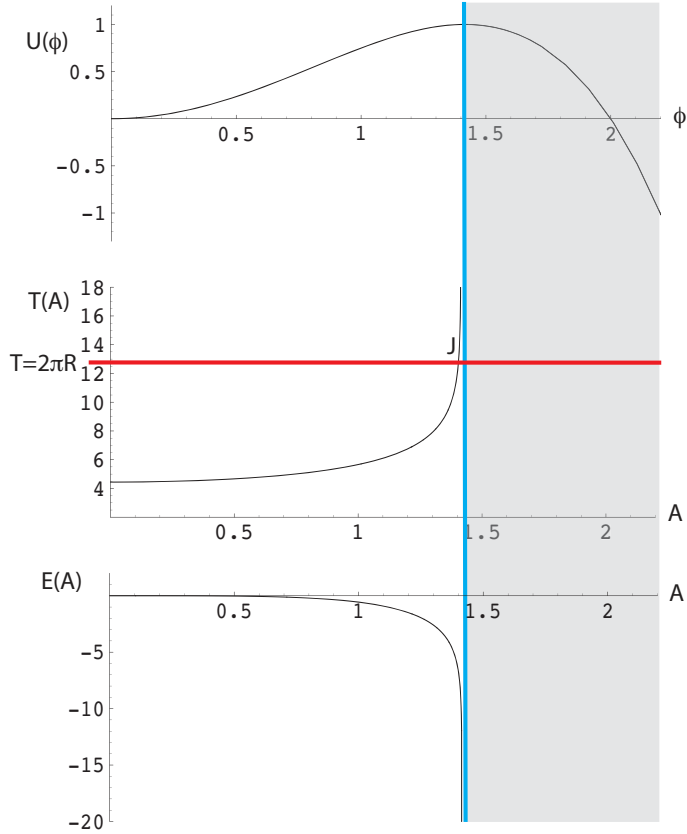}
\end{center}
\vspace{-.5cm}
\caption{For this example of the potential, given by~(\ref{v5}), we plot the inverted Potential $U_1(\phi)=-V_1(\phi)$, choosing
  $\mu^2=2$ and ${\bar \lambda} =1$ (top), the period function $T_1(A)$ (middle) and the energy $E_1(A)$ (bottom). There exists a
  unique, nodeless solutions, here labeled as $J$, which is stable. In the
  shaded regions there are no solutions with the appropriate boundary
  conditions.}
\label{example1plot}
\vspace{.2cm}
\end{figure}

It is easy to see that for $E={\mu^4\over 4{\bar \lambda}}$ one obtains non-trivial solutions known 
as the kink and anti-kink
\beq
\phi_{\rm (anti-)kink}(y)= \pm {\mu \over \sqrt{{\bar \lambda}}} \tanh\left[{\mu \over \sqrt{2}}\ (y-y_o)\right] \ ,
\label{kink}
\eeq
where the kink location $y_o$ should be set to zero because of the boundary
conditions of the scalar field.
This solution interpolates along the (now infinite) extra dimension between the
constant background solutions $\phi_{\pm}\equiv \pm{\mu/ \sqrt{{\bar \lambda}}}$.

For $0<E<{\mu^4\over 4{\bar \lambda}}$, we can still integrate~(\ref{energy_generalVA})
to obtain~\cite{Grzadkowski:2004mg}
\beq
\phi_k(y)=\pm 
{\mu\over \sqrt{{\bar \lambda}}} \sqrt{2 k^2 \over k^2+1}\   {\rm sn}\left({\mu\over
    \sqrt{k^2 +1}}\ y ,\ k^2 \right) \ ,
\label{periodicvev}
\eeq
where 
\beq
k^2
={\mu^2-\sqrt{\mu^4-4 {\bar \lambda} E} \over \mu^2+\sqrt{\mu^4-4 {\bar \lambda} E} }
\eeq
and ${\rm sn}(x,k^2)$ is the Jacobi Elliptic Sine-Amplitude, parametrized by the elliptic modulus $k$
(a real parameter such that $0<k<1$).
\begin{figure}[h]
\begin{center}
\includegraphics[width = 10cm]{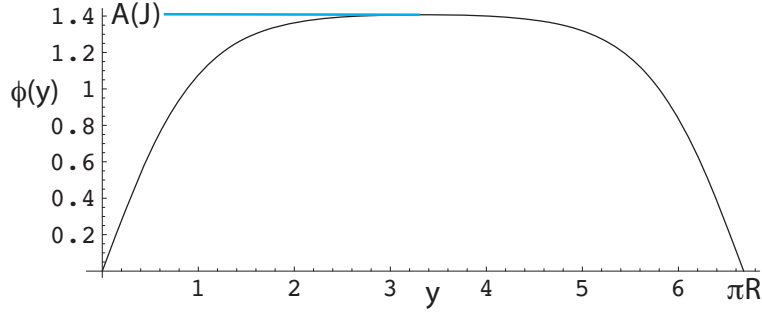}
\end{center}
\vspace{-.5cm}
\caption{The single stable solution (point $J$ in figure~\ref{example1plot}) for the potential~(\ref{v5}). 
Here we have chosen $T_1=2\pi R\simeq13.2$.}
\label{QSsolutionsEx1}
\vspace{.2cm}
\end{figure}
Its period is $4 K$, where
\bea
K(k^2)=\int_0^{\pi/2} {d\theta \over \sqrt{1 - k^2 \sin^2\theta}}
\eea
is the complete elliptic integral of the first kind.

While the above notation provides a natural way to think about this equation, it is convenient (and simple) to 
rewrite the solutions in terms of the amplitude parameter $A$, for consistency with the notation of the previous
sections, as
\beq
\phi_\subA(y)=A\  {\rm sn}\left(\sqrt{\mu^2+{{\bar \lambda}\over 2}A^2} y,\
  k^2\equiv {A^2\over 2{\mu^2\over{\bar \lambda}}+A^2 } \right) \ ,
\eeq
where the relationship between the amplitude $A$ and the constant of integration $E$ is 
\beq
A^2= {\mu^2-\sqrt{\mu^4-4{\bar \lambda} E}\over {\bar \lambda}} \ .
\eeq

In this example the total number of nontrivial solutions is given by 
$n_{\rm max}={\rm IP}(\mu R) -1$. Since $\mu$ is a fixed parameter of the scalar potential and $R$ is the fixed radius of the
extra dimension, $n_{\rm max}$ is completely specified by the model.

The complete set of static nontrivial background solutions consistent
with the boundary conditions, for the potential~(\ref{v5}) is then 
\beq
\phi_{k_n}(y)=\pm {\mu\over \sqrt{{\bar \lambda}}} \sqrt{2 k_n^2 \over k_n^2+1}\
  {\rm sn}\left({\mu \over \sqrt{k_n^2 +1}}\ y ,\ k_n^2 \right) \ ,
\eeq
where $n$ is an integer such that $\ 0 \le n \le n_{\rm max}$. 

The solution with lowest
energy, and no nodes in the interval, will be $\phi_{k_0}(y)$ (using $k$ as an equivalent label to $A$)
and is plotted in Fig.~(\ref{QSsolutionsEx1}). The rest of 
solutions $\phi_{k_n}(y)$ will 
have nodes and increasing energy. And thanks to our general stability argument
they will be unstable.

Note that the radius $R$ of the extra dimension is related to $k_0$ by
\beq
2\pi R={4\over \mu}\sqrt{k_0+1}\ K(k_0^2) \ .
\eeq

In~\cite{Grzadkowski:2004mg} the spectrum and eigenfunctions of the first few scalar excitations around
the nodeless background $\phi_{k_0}(y)$ were found. In our case the lowest-lying state is
\bea
 \varphi_0(y)&=&{\rm sn}\left(\sqrt{\mu^2+{{\bar \lambda}\over 2}A^2}\ y,\ k^2\!\equiv\!{A^2\over 2{\mu^2\over{\bar \lambda}}+A^2 }
\right)\non\\ 
&&\times\  {\rm dn}\left(\sqrt{\mu^2+{{\bar \lambda}\over 2}A^2}\ y,\
  k^2\!\equiv\!{A^2\over 2{\mu^2\over{\bar \lambda}}+A^2 }\right)\ \ ,
 \eea
where both ${\rm sn}$ and ${\rm dn}$ are Jacobi elliptic functions.
The mass eigenvalue of this lowest lying excitation is then given by 
\bea
\lambda\ \equiv\   M^2_0 \ = 
\ {3{\bar \lambda}\over 2} A^2. 
\eea
which is always positive, demonstrating, as expected, the stability of this solution.

\subsection{Example 2: Distorted Mexican Hat}

Even with just two degenerate minima, there exists the possibility for richer structure than in the simple model we have
just studied. To see this, consider a second potential
\beq
V_2(\phi)= -\phi^2 + {5\over 26} \phi^4 -{1\over 54} \phi^6 +{1\over 2000} \phi^8 \ ,
\label{secondexample}
\eeq
in which we have set all dimensionful parameters to unity.

\begin{figure}[h]
\begin{center}
\includegraphics[width = 9cm]{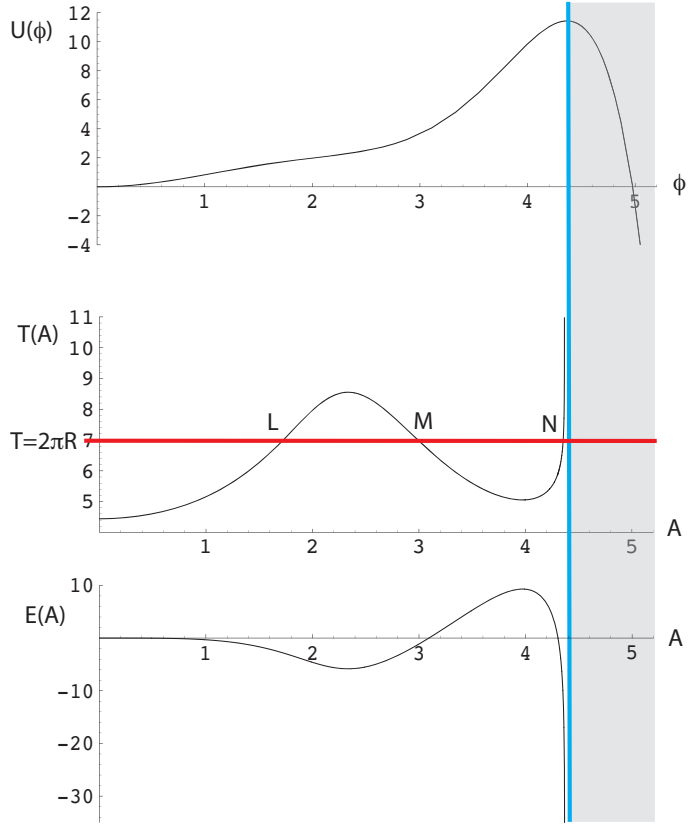}
\end{center}
\vspace{-.5cm}
\caption{For this example of the potential, given by~(\ref{secondexample}), we
  plot the inverted Potential $U_2(\phi)=-V_2(\phi)$ (top), the period
  function $T_2(A)$ (middle) and the energy $E_2(A)$ (bottom). There exist
  three distinct nodeless solutions, here labeled as $L$, $M$ and $N$, with
  different values of the amplitude $A$, but with the same period. The
  solution at $M$ is unstable, while those at $L$ and $N$ are stable. Further, by
  integrating~(\ref{EofA}) we find that $N$ is of lower energy than $L$. 
  In the shaded regions there are no solutions with the appropriate boundary conditions.}
\label{example2plot}
\vspace{.2cm}
\end{figure}

As in the previous example, this potential has only two degenerate minima, $\phi=\pm \phi_0$ at which
${\partial V_2\over\partial \phi}=0$ and
${\partial^2V_2\over\partial\phi^2}>0$. However, the crucial difference here
is that the second derivative of $V_2(\phi)$ vanishes at two additional
field values.

\begin{figure}[h]
\begin{center}
\includegraphics[width = 10cm]{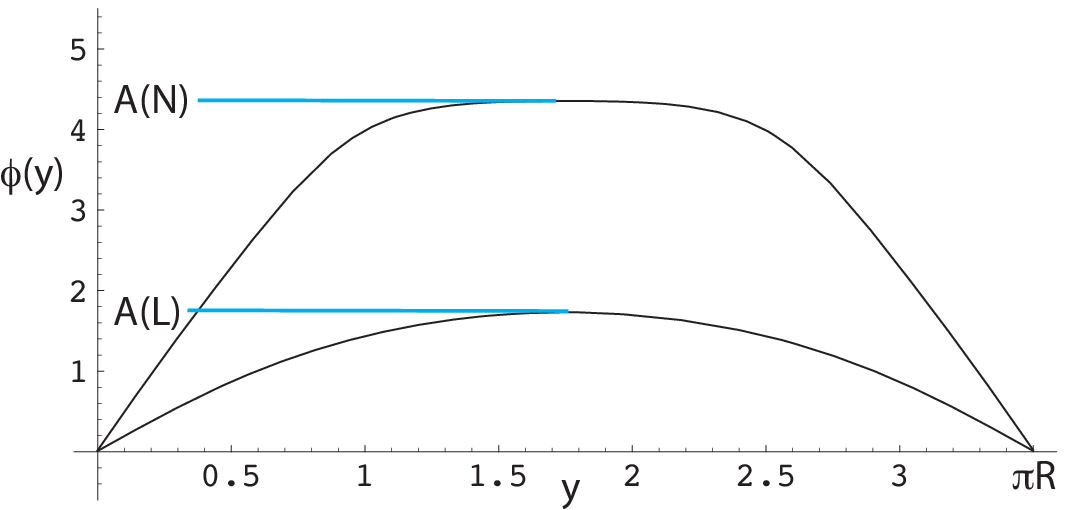}
\end{center}
\vspace{-.5cm}
\caption{The two stable solutions (points $L$ and $N$ in
  figure~\ref{example2plot}) for the potential~(\ref{secondexample}). The
  solution at point $N$, with the larger amplitude, $A(N)$, has the lower energy. Here we have chosen
  $T_2=2\pi R=7$.}
\label{QSsolutionsEx2}
\vspace{.2cm}
\end{figure}

This is enough to allow, for a certain range of choices of $\pi R$, the
existence of multiple nodeless solutions, illustrated by the points $L$, $M$ and $N$ on the
middle plot of figure~\ref{example2plot}. Our stability criterion then allows
us to immediately conclude that the solutions $L$ and $N$ are stable, while
solution $M$ is unstable. These stable solutions are shown in figure~\ref{QSsolutionsEx2}.

The bottom plot of figure~\ref{example2plot} represents the energy of
solutions as a function of amplitude and shows that the allowed nodeless
solution with higher amplitude ($N$ in this case) has the lower energy.

\subsection{Example 3: Many Local Minima}

Before concluding, let us provide a more complicated example
\beq
V_3(\phi)=-\phi^2 -5\phi^4+\frac{5}{2}\phi^6 -\frac{1}{3}\phi^8+\frac{1}{77}\phi^{10} \ ,
\label{thirdexample}
\eeq
in which we have set all dimensionful parameters to unity. This potential possesses a pair of degenerate local minima at
$\phi =\pm \phi_1$ and a distinct pair of degenerate global minima at $\phi=\pm \phi_2$.

\begin{figure}[t]
\includegraphics[width = 9cm]{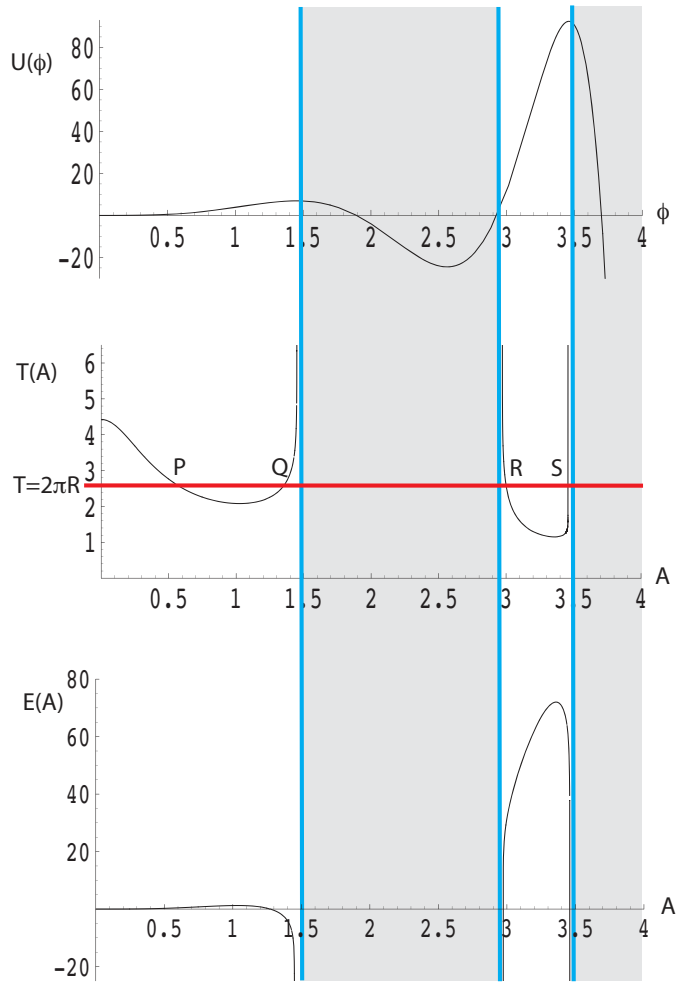}\\
\vspace{-.5cm}
\caption{For this example of the potential, given by~(\ref{thirdexample}), we
  plot the inverted Potential $U_3(\phi)=-V_3(\phi)$ (top), the period
  function $T_3(A)$ (middle) and the energy $E_3(A)$ (bottom). There exist
  four distinct nodeless solutions, here labeled as $P$, $Q$, $R$ and $S$, 
  with different values of the amplitude $A$, but with the same period. 
  Those at $P$ and $R$ are unstable, while those at $Q$ and $S$ are stable. 
  Integrating~(\ref{EofA}) we find that $S$ is of lower energy than $Q$. In the
  shaded regions there are no solutions with the appropriate boundary conditions.}
\label{example3plot}
\vspace{.2cm}
\end{figure}

\begin{figure}
\includegraphics[width = 10cm]{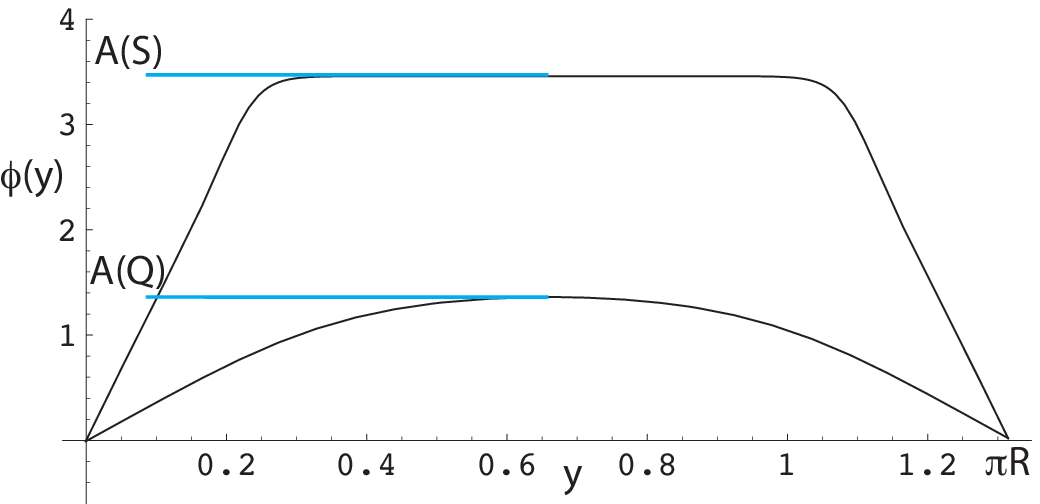}
\vspace{-.4cm}
\caption{The two stable solutions (points $Q$ and $S$ in
  figure~\ref{example3plot}) for the potential~(\ref{thirdexample}). The
  solution at point $S$, with the larger amplitude, $A(S)$, has the lower
  energy. Here we have chosen $T_3=2\pi R=2.6$.}
\label{QSsolutionsEx3}
\vspace{.2cm}
\end{figure}

In this case there are two separate intervals of the amplitude $A$ for which
there exist nontrivial solutions, as seen in Figure~\ref{example3plot}. With
the choice $2\pi R=2.6$, the middle plot shows that there are 4 nodeless
solutions $P$, $Q$, $R$ and $S$. Only two of them, $Q$ and $S$, shown in Figure~\ref{QSsolutionsEx3}, will be stable
according to our stability condition and we find that the larger amplitude
solution $S$ has the lower energy. 

\section{Conclusions and Outlook}

A thorough understanding of the implications of extra dimensional models
requires us to investigate not only perturbative phenomena, but also the
allowed distinct background configurations of brane and bulk fields. In
infinite dimensions, it is well-known that scalar fields with vacuum manifolds
with particular topological properties can give rise to topologically distinct
sectors of the theory, characterized by the soliton number. In a compact dimension, the
situation is more subtle, since the boundary conditions can affect the
stability of configurations identified in the infinite size limit.\\
\indent In this paper we have studied static, background configurations of scalar
fields in constructions in which the bulk space is an $S^1/Z_2$ orbifold - an interval with
reflection-symmetric boundary conditions. 
We have performed a general
stability analysis of such configurations, demonstrating that all solutions
with nodes in the interval are unstable. We have
also derived a powerful general criterion with which to determine the
conditions under which nodeless solutions are stable.\\
\indent In many cases, there are multiple nodeless solutions, in which case we 
need to determine which one is the vacuum state of the theory by computing its
associated energy density.\\
\indent The application of these results to model building and problem solving in 
extra dimension models may have novel and interesting implications for particle physics,
cosmology and the details of stabilization methods. To fully understand such
effects will require the inclusion of both quantum effects and gravity, a task
that is underway.

\acknowledgments
We thank Tim Tait and James Wells for discussions.
M.~Toharia is supported by funds provided by the University of Maryland,
Syracuse University and the U.S. Department of
Energy under Contract number. DE-FG-02-85ER 40231. 
M.~Trodden is supported by the National Science Foundation under 
grant PHY-0354990 and by Research Corporation.


\end{document}